\documentstyle[epsf]{mn}
\newcommand{\bm}[1]{{\mbox{\boldmath $#1$}}}
\def\ltsima{$\; \buildrel < \over \sim \;$}
\def\lsim{\lower.5ex\hbox{\ltsima}}
\def\gtsima{$\; \buildrel > \over \sim \;$}
\def\gsim{\lower.5ex\hbox{\gtsima}}

\twocolumn
\title{Substructure and the halo model of large-scale structure}

\author[R. K. Sheth \& B. Jain]{Ravi K. Sheth$^1$ \& Bhuvnesh Jain$^2$\\
$^1$ Department of Physics and Astronomy, University of Pittsburgh, 
     3941 O'Hara Street, Pittsburgh, PA 15260\\
$^2$ Department of Physics and Astronomy, University of Pennsylvania, 
     209 S. 33 Street, Philadelphia, PA 19104}

\begin{document}

\maketitle

\begin{abstract}
We develop the formalism to include substructure in the halo 
model of clustering. Real halos are not likely to be 
perfectly smooth, but have substructure which has so far been
neglected in the halo model --- our formalism
allows one to estimate the effects of this substructure on measures of 
clustering. We derive expressions for the two-point correlation function, 
the power-spectrum, the cross-correlation between galaxies and 
mass, as well as higher order clustering measures.  Simple 
forms of the formulae are obtained for the limit in which the size of
the substructure and mass fraction in it is small. Inclusion 
of substructure allows for a more accurate analysis of the statistical 
effects of gravitational lensing.  It can also bring the halo 
model predictions into better agreement with the small-scale 
structure seen in recent high resolution simulations of hierarchical 
clustering.  
\end{abstract}


\section{Introduction}

Sheth \& Jain (1997) described how the halo model for clustering
can allow one to model 
the distribution of matter in the highly nonlinear regime.  The model 
falls within the 
broader framework described by Neyman \& Scott (1954) and 
Scherrer \& Bertschinger (1991).  It combines results from Peebles (1974) 
and McClelland \& Silk (1977) with the work of Press \& Schechter (1974), 
and is able to provide a good description of nonlinear clustering seen in 
numerical simulations of hierarchical gravitational clustering.  
The halo model assumes that most of the mass in the Universe is bound up in 
virialized dark matter halos, and that statistical measures of clustering 
on small scales are dominated by the internal structure of the halos.  
The agreement with simulations shows that is possible to provide an 
accurate description of clustering in the small-scale nonlinear regime 
even if one has no knowledge of if and how the halos themselves are 
clustered.  This halo--model of clustering has been the subject of much 
recent interest (e.g. Seljak 2000; Peacock \& Smith 2000; Ma \& Fry 2000; 
Scoccimarro et al. 2001; Cooray \& Sheth 2002).  

To date, almost all analytic work based on the halo--model approach 
assumes that halos are spherically symmetric, and that the density 
run around each halo center is smooth.  Halos which form in numerical 
simulations of hierarchical clustering are neither spherically symmetric 
nor smooth (e.g., Navarro, Frenk \& White 1997; Moore et al. 1999; 
Jing \& Suto 2002).  
About ten percent of the mass of a halo is associated with subclumps 
(Tormen, Diaferio \& Syer 1998; Ghigna et al. 1999).  
The main purpose of the present work is to derive a model which 
accounts for this substructure.  

Section~\ref{themodel} shows how to compute two-point statistics, the 
correlation function and its Fourier transform, the power spectrum,
when substructure is important.  Section~\ref{higher} shows that the 
model can be easily extended to estimate higher-order statistics.  
Section~\ref{averages} shows how our formalism can incorporate a 
range of parent halo and subclump masses, and Section~\ref{details} 
provides a few explicit examples.  This section includes a discussion 
of how to model the shape of the subclump mass function.  
Section~\ref{discuss} summarizes our results, and suggests various other 
applications of our formalism.  

\section{Two-point correlations}\label{themodel}
The halo model approach assumes that all mass is bound up in dark 
matter halos.  All statistical measures of clustering are then decomposed 
into two distinct types of contributions: one comes from sets of particles 
which are in the same halo, and the other comes from particles which are 
in different halos.  The assumption is that, on the smallest scales,
it is the single-halo contribution which dominates the statistic.  
In the context of the present paper, this means that we expect that the 
contribution to, e.g., the power spectrum, which comes from substructure 
cannot be important on scales larger than a typical halo.  Therefore, even 
if substructure changes the single-halo contribution substantially compared 
to the case of smooth halos, it should be accurate to ignore substructure 
when estimating the two- and higher-order halo terms which dominate on 
larger scales.  The primary consequence of substructure, then, is that 
the single-halo term becomes more complicated.  For example, for two-point 
statistics, the one-halo term will now have three types of pairs:  
both particles from the smooth component, 
both particles from the substructure component, 
and one particle from the smooth component with the other from the 
substructure.  Our goal will be to derive expressions for these 
different contributions.  

\subsection{Preliminaries}
The probability of finding a subclump in the volume element 
$d^3{\bm r}$ at distance ${\bm r}$ from the center of the parent halo 
of mass $M$ is
\begin{equation}
 p({\bm r})\,d^3{\bm r} = 
   {n_c({\bm r})\,d^3{\bm r}\over \int d^3{\bm r}\,n_c({\bm r})}.  
\end{equation}
Here $n_c$ describes the number density of subclumps at ${\bm r}$.  
In what follows, it will prove convenient to multiply this number 
density by a mass and so define a mass density $\rho_c$.  We choose 
this mass so that $\int d^3{\bm r}\,\rho_c({\bm r}) = M$, the total 
mass in the halo.  
(For simplicity we will sometimes assume that the subclumps have the 
same density run as the smooth dark matter distribution, but our 
analysis is not confined to this case.  In this special case, $\rho_c$ 
is the same as the mass density profile of the smooth component.)  

The density at ${\bm r}$ from the halo center is the sum of the smoothly 
distributed mass plus the contribution from the subclumps.  If $F$ 
and $f_i$ denote the density run in units of the mean density $\bar\rho$ 
around the parent halo and the $i$th subclump respectively, then 
\begin{equation}
 {\rho({\bm r})\over \bar\rho} = F({\bm r}) 
                                 + \sum_{i=1}^N f_i({\bm r}-{\bm r}_i).
 \label{rhoexpansion}
\end{equation} 
The mass in the smooth and subclump components is defined by 
\begin{equation}
 M_s \equiv \int dr\,4\pi r^2\,\bar\rho\ F(r) \quad {\rm and}\quad 
 m_i \equiv \int dr\,4\pi r^2\,\bar\rho\ f(r).
\end{equation}
The total mass $M$ is the sum of the smooth and clumped components.  

Our strategy will be to first derive expressions for, e.g., the 
correlation function for fixed values of $M$, $m_i$ and $N$, and to 
average over the distributions of these variables later.  We will keep
the discussion as general as possible; realistic choices for $F$, $f$,
$p$, and the halo and subclump mass functions will be inserted into 
the formalism later.  

\subsection{The correlation function of the mass}

The ensemble-averaged two-point correlation function $\xi(r)$ will be
obtained by averaging over halos with distributions of sub-structure specified
by $p(r)$. We will first obtain the two-point correlation function within
a halo, denoted ${\cal C}({\bm r})$, then average over substrucutre to
get $\xi(r)$, and finally in section 4 consider the averaging over 
varying numbers of subclumps and of parent halo masses. For a halo of
total mass $M$ we define
\begin{equation}
 {\cal C}({\bm r}) \equiv \bar n
   \int d^3{\bm s}\ \delta({\bm s})\delta({\bm s}+{\bm r}),\ \  
   {\rm where} \ \ \delta({\bm r}) \equiv {\rho({\bm r})\over\bar\rho} - 1,
\end{equation}
and $\bar n$ denotes the number density of the parent halos.  
If the mass densities within the subclumps are all much larger than the 
mean density $\bar\rho$, then 
$\delta({\bm r}) \approx \rho({\bm r})/\bar\rho$.  
Writing out all the terms explicitly shows that
\begin{eqnarray}
 {{\cal C}({\bm r})\over\bar n} &=& 
  \int d^3{\bm s}\ F({\bm s})\,F({\bm s}+{\bm r})  \nonumber\\
 && + \sum_i \int d^3{\bm s}\ f_i({\bm s}-{\bm r}_i)\,
                             f_i({\bm s}+{\bm r}-{\bm r}_i)   \nonumber\\
 && + \sum_j \sum_{i\ne j} \int d^3{\bm s}\ f_i({\bm s}-{\bm r}_i)\,
                                 f_j({\bm s}+{\bm r}-{\bm r}_j) \nonumber \\
 && + \sum_i \int d^3{\bm s}\ F({\bm s}+{\bm r})\,f_i({\bm s}-{\bm r}_i) 
      \nonumber \\
 && + \sum_i \int d^3{\bm s}\ F({\bm s})\,f_i({\bm s}+{\bm r}-{\bm r}_i).
 \label{lambdar}
\end{eqnarray}
This is the correlation function associated with a smooth component and 
the $i$ subclumps at the specified positions.  
The first term is the contribution from particles which are not in the 
subclumps, and is familiar from previous work; this is the contribution 
from the first pair-type.  
The second term is from pairs where both particles are in the same 
subclump, whereas the third term is from pairs where the two particles 
are in separate subclumps.  These two terms represent the contribution 
from the second pair-type.  The fourth and fifth terms are from pairs in 
which one particle is in a subclump and the other is in the smoother 
component.  

We are less interested in the correlation function associated with a 
specific realization of the subclump distribution, than we are with what 
happens upon averaging over the various possible subclump distributions.  
That is, we are more interested in 
\begin{equation}
 \xi({\bm r}) = \int d^3{\bm r}_i\int d^3{\bm r}_j \ 
                    p( {\bm r}_i)\,p( {\bm r}_j)\ {\cal C}({\bm r}).
\label{xir}
\end{equation}
The next step is to compute these averages.   

Let $\lambda_{ss}({\bm r})$ denote the value of the first integral 
in expression~(\ref{lambdar}).  It is the same for all the possible 
subclump distributions, so the result of averaging it over the distributions 
is 
\begin{equation}
 \lambda_{ss}({\bm r}) = 
          \int d^3{\bm s}\,F({\bm s})\,F({\bm s}+{\bm r}).
\end{equation}
This term is the convolution of $F$ with itself, and is the only 
term which most halo--models use.  The remaining terms are due to 
the substructure.  Before we compute them, it will prove convenient 
to rewrite this first term as 
\begin{displaymath}
 \lambda_{ss}({\bm r}) = 
 \left(M_{s}\over M\right)^2 \lambda_{\rm smooth}({\bm r}) =
 \left(1 - \sum_i {m_i\over M}\right)^2 \lambda_{\rm smooth}({\bm r}).
\end{displaymath}
Here $\lambda_{\rm smooth}({\bm r})$ denotes the correlation function 
if the total mass were smoothly distributed around the center of the 
parent halo (i.e, if there were no subclumps).  Thus, the factor of 
$(M_{s}/M)^2$ simply denotes the fact that now only a fraction of 
the mass is in the smooth component.  

Let $\lambda_{ii}({\bm r})$ denote the result of averaging each of the 
second terms in equation~(\ref{lambdar}) over $p({\bm r}_i)$.  Then 
\begin{eqnarray}
 \lambda_{ii}({\bm r}) &=& \int d^3{\bm r}_i\,p({\bm r}_i)
                     \int d^3{\bm s}\ f_i({\bm s}-{\bm r}_i)\,
                                      f_i({\bm s}+{\bm r}-{\bm r}_i)\nonumber\\
                    &=& \int d^3{\bm x}\ f_i({\bm x})\,f_i({\bm x}+{\bm r}),
\end{eqnarray}
where we have set ${\bm x}={\bm s}-{\bm r}_i$.  This term is the 
convolution of each subclump profile $f_i$ with itself.  

Averaging the third term in equation~(\ref{lambdar}) is more complicated 
so we will do it last.  The result of averaging the fourth term is 
\begin{eqnarray}
 \lambda_{si}({\bm r}) &=& \int d^3{\bm r}_i\,p({\bm r}_i) 
       \int d^3{\bm s}\ F({\bm s}+{\bm r})\,f_i({\bm s}-{\bm r}_i)\nonumber\\
  &=& \int d^3{\bm s}\int d^3{\bm r}_i\ f_i({\bm s}-{\bm r}_i)\ 
  F({\bm s}+{\bm r}) 
  {\rho_c({\bm r}_i)\over \int d^3{\bm r}\,\rho_c({\bm r})} \nonumber\\
  &=& \int d^3{\bm x}\ f_i({\bm x}) \int d^3{\bm r}_i\  
  F({\bm x}+{\bm r_i}+{\bm r})\, {\rho_c({\bm r}_i)\over M} \nonumber\\
  &=& \int {d^3{\bm x}\over  M/\bar\rho}\  f_i({\bm x})\, 
      \lambda_{sc}({\bm x}+{\bm r})
\end{eqnarray}
where $\lambda_{sc}$ denotes the convolution of $F$ with 
$\rho_c/\bar\rho$.  
(If the subclumps were distributed around the halo center similarly to 
the dark matter, i.e., $\rho_c/\bar\rho\propto F$, then 
$\lambda_{sc}\propto\lambda_{ss}$.)  
By symmetry, the average of the fifth term is the same.  

And finally, the average of the third term is 
\begin{eqnarray} 
 \lambda_{ij}({\bm r}) &=& \int d^3{\bm r}_i\int d^3{\bm r}_j\ 
                           p({\bm r}_i)\,p( {\bm r}_j)\ \nonumber\\
  && \qquad \times         \int d^3{\bm s}\ f_i({\bm s}-{\bm r}_i)\,
                                f_j({\bm s}+{\bm r}-{\bm r}_j) , 
\end{eqnarray}
where the integral over ${\bm s}$ is the convolution of the two subclump
profiles. 
Written this way, $\lambda_{ij}$ is the weighted sum of this convolution
over all pairs of subclump 
positions. 
It is interesting to re-arrange the order of the 
integrals above:
\begin{eqnarray} 
 \lambda_{ij}({\bm r}) &=& \int d^3{\bm r}_i\int d^3{\bm r}_j\ 
                       p({\bm r}_i)\,p( {\bm r}_j)\ \nonumber\\
     &&\qquad \times     \int d^3{\bm s}\ f_i({\bm s}-{\bm r}_i)\,
                            f_j({\bm s}+{\bm r}-{\bm r}_j) \nonumber\\
   &=& \, \int d^3{\bm s} \int d^3{\bm r}_i\ f_i({\bm s}-{\bm r}_i)\nonumber\\
     && \, \, \times \int d^3{\bm r}_j f_j({\bm s}-{\bm r}_i+{\bm r}+{\bm r}_{ij}) 
             p({\bm r}_{ij}+{\bm r}_j)\,p( {\bm r}_j) \nonumber\\
 &=& \int d^3{\bm x} \int d^3{\bm r}_{ij}\ f_i({\bm x}) 
    \,f_j({\bm x} +{\bm r} + {\bm r}_{ij})\,
     {\lambda_{cc}({\bm r}_{ij})\over (M/\bar\rho)^2} 
     \nonumber\\
 &=& \int d^3{\bm x} \int d^3{\bm y}\ f_i({\bm x})\,f_j({\bm y})\,
     {\lambda_{cc}({\bm y}-{\bm x}-{\bm r}) \over (M/\bar\rho)^2} 
\end{eqnarray}
where $\lambda_{cc}$ denotes the convolution of $\rho_c/\bar\rho$ 
with itself.  The second line follows from setting 
${\bm r}_{ij} = {\bm r}_i - {\bm r}_j$, the third line from setting 
${\bm x}={\bm s}-{\bm r}_i$ and the final expression from setting 
${\bm y}={\bm x} +{\bm r} + {\bm r}_{ij}$.  When written in this way, 
$\lambda_{ij}$ appears very similar to the two-halo term when substructure 
is absent; $\lambda_{cc}$ plays the role of the `subclump--subclump 
correlation function' (compare equation~2 in Sheth et al. 2001).  

Using the above results, 
the correlation function $\xi(r)$ defined in equation (\ref{xir})
can be expressed as
\begin{equation}
 {\xi({\bm r})\over\bar n} = \lambda_{ss}({\bm r}) 
                   + \sum_i \lambda_{ii}({\bm r})
                   + 2\sum_i \lambda_{si}({\bm r})
                   + \sum_i\sum_{j\ne i} \lambda_{ij}({\bm r}).  
\end{equation}
Note that all the $\lambda$ terms on the right-hand side have dimensions
of volume, so that $\xi$ is dimensionless. 

\subsection{Limiting cases for small-sized subclumps}

Now consider some simple limiting cases.  
If the subclumps are much smaller than the parent halo, then we can 
treat them as point masses when performing the integrals which define 
$\lambda_{si}$ and $\lambda_{ij}$.  The associated delta function 
profiles simplify the integrals, so that 
\begin{eqnarray}
 {\xi({\bm r})\over\bar n} &\approx & \lambda_{ss}({\bm r}) 
       + \sum_i \lambda_{ii}({\bm r})
       + 2\sum_i {m_i\over M}\,\lambda_{sc}({\bm r}) \nonumber\\
 && \qquad\quad + \sum_i\sum_{i\ne j} {m_i\over M}{m_j\over M}\,\lambda_{cc}({\bm r}).
\end{eqnarray}
If in addition, the distribution of subclumps around the halo center 
is similar to the dark matter, $\rho_c\propto F$, then we can set 
$\lambda_{sc}\propto \lambda_{ss}$ and 
$\lambda_{cc}\propto \lambda_{ss}$.  In this case, the previous 
expression becomes 
\begin{eqnarray}
 {\xi({\bm r})\over\bar n} &\to & \lambda_{ss}({\bm r})
  \left[ 1 + 2\sum_i {m_i\over M_{s}} 
           + \sum_i\sum_{j\ne i} {m_im_j\over M_{s}^2}\right] \nonumber\\
 && + \sum_i \lambda_{ii}({\bm r})\nonumber\\
 &=& \lambda_{\rm smooth}({\bm r}) 
    \left[ 1 - \sum_i \left({m_i\over M}\right)^2\right]
    + \sum_i \lambda_{ii}({\bm r}).
 \label{xiapprox}
\end{eqnarray}
(Recall that $\lambda_{\rm smooth}$ denotes the correlation function
if all the mass was smoothly distributed around the halo center.)  
If the fraction of the total mass which is in subclumps is small, then 
$\xi({\bm r})\approx \bar n \lambda_{\rm smooth}({\bm r}) 
                     + \sum_i \bar n \lambda_{ii}({\bm r})$.  
In this limit, the total correlation function is well approximated by 
taking what one would have got if the mass was smoothly distributed, 
and then adding the contribution from the individual subclump components.  
In the small mass and size limit, the contribution from the subclumps can 
only be important on scales smaller than the typical subclump, so that 
this additional contribution is only important on very small scales.  
On scales larger than the diameter of a typical subclump, the correlation 
function looks just as though the mass within halos is smoothly 
distributed.  This simple assumption is probably sufficiently accurate 
for most applications.  


We remarked that the term $\lambda_{cc}({\bm r})$ could be thought of 
as the correlation function of the subclumps.  To see why, suppose 
there is no mass in the smooth component, and all the subclumps are 
infinitesimally small:  i.e., we replace all factors of the subclump 
density profile $f_i$ with delta functions $\delta_{\rm D}({\bm r}_i)$.  
Then, when ${\bm r}\ne 0$, only the third term in equation~(\ref{lambdar}) 
contributes any pairs.  Using the delta functions reduces this term to 
\begin{eqnarray}
 \lambda_{ij}({\bm r})  &=& \sum_i \sum_{j\ne i} 
  \int d^3{\bm r}_i\int d^3{\bm r}_j\ p({\bm r}_i)\,p( {\bm r}_j)\nonumber\\
 &&\qquad\times
  \int d^3{\bm s}\ f_i({\bm s}-{\bm r}_i)\,f_j({\bm s}+{\bm r}-{\bm r}_j)
   \nonumber\\
 &=& \sum_i\sum_{j\ne i} 
\left(m_i m_j\over\bar\rho^2\right)\ \int d^3{\bm r}_i\ 
           p({\bm r}_i)\,p({\bm r}_i+{\bm r}),
\end{eqnarray}
which is proportional to the convolution of the subclump distribution 
$\rho_c$ with itself.  

\subsection{The cross--correlation between subclumps and mass}

We can use a similar argument to compute the cross correlation between 
subclumps and mass.  That is, we imagine we sit at the center of the 
$i$th subclump and we then compute the typical density at distance 
${\bm r}$ from it.  Because the first term in equation~(\ref{lambdar}) 
is not centered on a subclump it no longer contributes.  Since our 
constraint only requires one member of each pair of positions to be 
centered on a subclump we now need to replace one factor of $f_i$ 
with a delta function.  This means 
\begin{eqnarray*}
 {\bar n \ \lambda_{ii}({\bm r})} &\to & f_i({\bm r}), \quad 
 {\bar n \ \lambda_{si}} \to 
     {\lambda_{sc}({\bm r})\over (M/\bar\rho)},
 \quad{\rm and}\quad \nonumber\\
 {\bar n \ \lambda_{ij}({\bm r})} &\to &
  \int d^3{\bm y}\,f_j({\bm y}){\lambda_{cc}({\bm y-r})\over (M/\bar\rho)^2}, 
\end{eqnarray*}
so that the subclump--mass cross correlation function is 
\begin{eqnarray}
 {\xi_\times({\bm r})} &=& 
                      \sum_i {\lambda_{sc}({\bm r})\over M/\bar\rho} 
                       + \sum_i f_i(\bm r) \nonumber\\  
   &&\qquad            + \sum_i \sum_{j\ne i}\int d^3{\bm s}\ f_j({\bm s})\, 
                       {\lambda_{cc}({\bm s} - {\bm r})\over (M/\bar\rho)^2}.
\end{eqnarray}
In the limit in which the subclumps are much smaller in size than
the parent halo the last term simplifies to give, 
\begin{displaymath}
 {\xi_\times({\bm r})}\approx 
 \sum_i {\lambda_{sc}({\bm r})\over M/\bar\rho} 
                         + \sum_i f_i(\bm r)  
                         + \sum_i\sum_{j\ne i} {m_j\over M}\, 
                         {\lambda_{cc}({\bm r})\over (M/\bar\rho)}.
\end{displaymath}
When $\rho_c/\bar\rho\propto F$ this becomes 
\begin{equation}
 \xi_\times({\bm r}) \approx 
 \sum_i {\lambda_{ss}({\bm r})\over M/\bar\rho}
  \left[1 + \sum_{j\ne i} {m_j\over M}\right] + \sum_i f_i(\bm r), 
\end{equation}
and further, if the  total mass in subclumps is small we obtain
\begin{equation}
 \xi_\times({\bm r})\approx \sum_i\lambda_{ss}({\bm r})/(M/\bar\rho) 
   + \sum_if_i(\bm r) .
\end{equation}
It is easy to verify that the result for $\xi_\times$ can 
be obtained simply by averaging 
\begin{eqnarray}
 \sum_i \rho_{cm|i}({\bm r}_i + {\bm r}) &=& 
 \sum_i F({\bm r}_i + {\bm r}) + \sum_i f_i({\bm r}) \nonumber\\ 
  &&\quad          + \sum_i \sum_{j\ne i} f_j({\bm r}_i + {\bm r}-{\bm r}_j)
\end{eqnarray}
over all realizations of subclump positions, $p({\bm r}_i)$ and 
$p({\bm r}_j)$.  

\subsection{Power spectra}
The power spectrum is the Fourier Transform of the correlation function.  
Since the correlation function involves a number of convolution-type 
integrals, the power spectrum is given by simple multiplications of 
the various density profile factors.  Let 
\begin{equation}
 U(k) = {\int dr\,4\pi r^2\,F(r)\,\sin(kr)/kr\over 
         (2\pi)^3\,\int dr\,4\pi r^2\,F(r)}
\end{equation}
denote the Fourier transform of the density run of the smooth component 
$F$ normalized by the total mass contained in the profile, and define 
$u_i(k)$ and $U_c(k)$ similarly (we have assumed we are working in 
three-dimensions).  Then 
\begin{eqnarray}
 P(k)\!\!\! &=&\!\!\! \int {dr\,4\pi r^2\over (2\pi)^3}\,{\sin(kr)\over kr} \ 
                \xi(r)\nonumber\\
     \!\!\! &=&\!\!\! \bar n\ \left({M\over\bar\rho}\right)^2\,
    \Biggl[\left({M_{s}\over M}\right)^2 U(k)^2 \nonumber\\
      & &\qquad\qquad\qquad  
           + 2\sum_i {m_iM_{s}\over M^2}\,u_i(k)\,U(k)\,U_c(k)\nonumber\\
      & &\qquad\qquad\qquad 
          + \sum_{i}\sum_{j\ne i} {m_im_j\over M^2}u_i(k)\,u_j(k)\,U_c(k)^2
      \nonumber\\
      & &\qquad\qquad\qquad  
          + \sum_i \left({m_i\over M}\right)^2 \,u_i(k)^2 \Biggr],
 \label{pk}
\end{eqnarray}
which we could write as 
\begin{displaymath}
 P(k)\equiv P_{ss}(k) + 2\sum_i P_{si}(k) 
               + \sum_i\sum_{j\ne i}P_{ij}(k) + \sum_i P_{ii}(k).
\end{displaymath}
A similar calculation shows that the cross spectrum between subclumps 
and mass is 
\begin{eqnarray}
 P_\times(k) &=& \ \sum_i 
               \Biggl[{M_{s}\over\bar\rho}\,U(k)\,U_c(k) \nonumber\\
           &&  \ \ +\ U_c(k)^2\sum_{j\ne i} {m_j\over \bar\rho}\,u_j(k)
                 + {m_i\over \bar\rho}\,u_i(k) \Biggr], 
\end{eqnarray}
and the power spectrum of the subclumps is 
\begin{equation}
 P_{cc}(k) = \bar n\ \left({M\over\bar\rho}\right)^2
                U_c(k)^2 \sum_i\sum_{j\ne i} {m_i\over M}{m_j\over M}.
\end{equation}

If we set $U_c(k)=U(k)$, and further assume that 
the subclumps are much smaller than the parent halo, 
then the power 
spectrum reduces to the sum of the power spectra of the individual 
components:  
\begin{eqnarray}
{P(k)\over \bar n} 
&\approx& \left(M\over \bar\rho\right)^2\ \left[ 1 - \sum_i \left({m_i\over M}
\right)^2\right]U(k)^2 \nonumber \\
&& + \ \sum_i \ \left(m_i\over \bar\rho\right)^2\,u_i(k)^2.
\label{powerapprox}
\end{eqnarray}
This shows that whether or not the substructure component dominates the 
small scale power depends on the fraction of mass in the subclumps, and 
on how much more dense the subclumps are compared to the smooth component.  

\begin{figure*}
\centering 
\epsfxsize=\hsize\epsffile{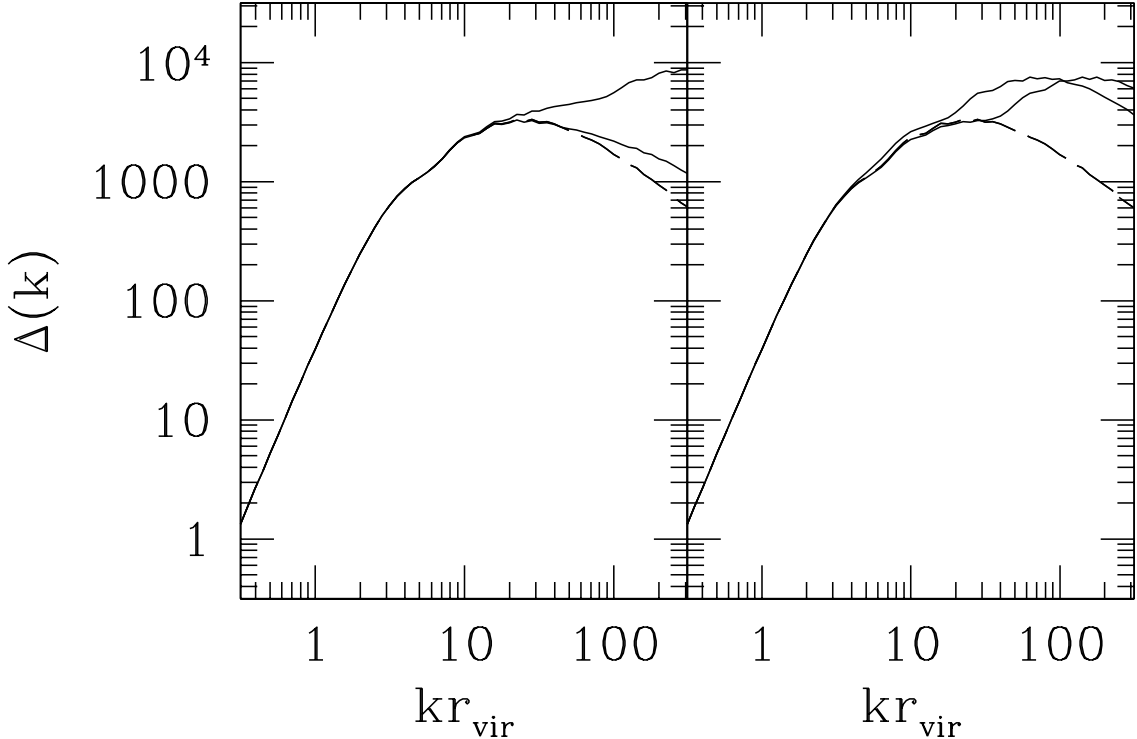}
\vspace{-6cm}
\caption{Effect on the power-spectrum as the parameters which 
describe the subclump distribution are varied.  The dashed line, 
which is the same in both panels, shows $\Delta(k)$ when all the mass 
is in smooth NFW halos of radius $r_{\rm vir}$ within which the 
average density is 200 times the background density.  The solid curves 
in the panel on the left show $\Delta(k)$ when $f=0.1$ of the total mass 
of each parent halo is in subclumps, each containing 1\% of the mass of 
the parent.  The average density within each subclump is $500\bar\rho$ 
(lower) and $5000\bar\rho$ (higher).  
In the panel on the right, $f=0.9$, the subclumps are 500 times denser 
than the background, and the subclumps are each 0.01 (peak at larger 
$k$) and 0.1 (peak at smaller $k$) times the mass of their parents.  }
\label{pksubclumps}
\end{figure*}

To see this a little more clearly, consider a specific example.  Suppose 
that the density runs are Gaussians, and that all the $N$ subclumps have 
the same mass $m_i$ and size $R_i$.   If $R$ denotes the characteristic 
scale of the density runs of the smooth component, then 
$U=\exp(-k^2R^2/2)$ and $u_i = \exp(-k^2R_i^2/2)$.  The subclumps 
dominate the power if $N(m_i/\bar\rho)^2 u_i^2 > (M/\bar\rho)^2\,U(k)^2$.
This happens on scales where 
$k^2 > [2\ln (M/M_{cl})+\ln N]/R^2/[1 - (R_i/R)^2]$, where we have 
set $M_{cl} \equiv Nm_i$.  Note that if the total mass in subclumps 
is small, $M_{cl}/M\ll 1$, then the subclumps dominate the power only 
at very large $k$; if this mass is divided up among many subclumps, 
the subclumps only dominate at even higher $k$.   For a fixed mass ratio, 
the exact scale on which the subclumps dominate depends on the size 
ratio $R_i/R$.  This suggests that a feature in $P(k)$ at large $k$ may 
provide information about the nature of the subclumps.  Simulations 
suggest that $M/M_{cl}\sim 10$ and $R_i\lsim R/10$, which yields a critical 
value of $k \gsim \sqrt{6 + \ln N}/R$.  It is interesting that this is 
just beyond the reach of current simulations.  

Fig.~\ref{pksubclumps} illustrates this with slightly more realistic 
Navarro, Frenk \& White (1997) density profiles.  The various curves show 
$\Delta(k)\equiv 4\pi k^3\,P(k)$ for models in which the density field
is made up of Poisson distributed NFW halos.  The parent halos are 
truncated at their virial radii $r_{\rm vir}$ defined so that the average 
density within $r_{\rm vir}$ is 200 times that of the background.  
We set the NFW core radius to be $a_{\rm NFW} =0.1\,r_{\rm vir}$.  
The dashed line (same in both panels) shows $\Delta(k)$ if there is 
no substructure.  
We then assumed that a fraction $f$ of the mass of each parent halo 
was in subclumps, each of mass $m$.  For simplicity, we assumed that 
the distribution of subclumps around the halo center was given by the 
same NFW form, and that the distribution of particles around each 
subclump center was also NFW, with core radius 
$a_{\rm NFW}=0.1\,r_{\rm sc}$.  The value of $r_{\rm sc}$ was set by 
requiring that the average density within the subclumps equal 
$\delta_{\rm sc}$ times the background density. 
The solid curves in the panel on the left show equation~(\ref{pk}) 
when $f=0.1$, $m=0.01M$ and $\delta_{\rm sc}=500$ (lower) and 
$\delta_{\rm sc}=5000$ (upper).  This shows that, all other things 
being equal, denser subclumps contribute more power.  
The solid curves in the panel on the right show results when 
$f=0.9$, $\delta_{\rm sc}=500$ and $m/M=0.1$ and $m/M=0.01$ 
(departure from dashed curve apparent at high and still higher $k$).  
Comparison with the panel on the left shows that increasing the 
fraction of mass in subclumps, $f$, increases the small scale power.  
The scale on which this increase becomes apparent depends on  
$(m/M)^{1/3}(\delta_{\rm vir}/\delta_{\rm sc})^{1/3}$, the 
typical radii of the subclumps.  

\section{Higher-order statistics}\label{higher}
Higher order correlations at a given small scale are dominated more 
strongly by the one-halo term than is the two point function, so the 
effect of substructure on the one-halo term is of great interest for 
higher order correlations. 
In this section we will consider the effect of substructure on 
the 3-point correlation function. The extension to higher orders is 
obvious.  

We begin with the expansion of the density $\rho(r)$ given in 
equation~(\ref{rhoexpansion}). 
The 3-point function in real space is then defined by
\begin{equation}
 \xi_{123}\equiv \left< \delta({\bm r_1}) \delta({\bm r_2})
                                           \delta({\bm r_3}) \right> \simeq
               {1\over \bar\rho^3} 
               \left< \rho({\bm r_1}) \rho({\bm r_2})\rho({\bm r_3}) \right> .
\label{3ptdefinition}
\end{equation}
It is convenient to consider the expressions for higher order
correlations in Fourier space, since we will find that in the
limiting case of interest the only additional term due to 
substructure is a term involving products of the substructure profile. 

The Fourier transform of the 3-point function is the bispectrum
$B$ defined by
\begin{equation}
 \Bigl<{\delta(\bm{k}_1)\delta(\bm{k}_2)\delta(\bm{k}_3)}\Bigr>
 = B(\bm{k}_1,\bm{k}_2,\bm{k}_3) \ (2\pi)^3 \ \delta_D(\bm{k}_{123}), 
\label{bispectrum}
\end{equation}
where the Delta function indicates that ${\bm k_1 + \bm k_2 + \bm k_3}=0$.  

It is the sum of contributions from triplets which are all in the 
smooth component, triplets all in the same halo, plus contributions 
from the various cross terms.  Writing all the different terms 
explicitly gives 
\begin{eqnarray}
&& B(k_1,k_2,k_3) = \bar n  
    \biggl[\left({M_s\over\bar\rho}\right)^3 U(k_1) U(k_2) U(k_3) \nonumber\\
  && + \sum_i {m_i\over\bar\rho} u_i(k_1) 
  \left({M_s\over\bar\rho}\right)^2 U(k_2) U(k_3)\,U_c(k_1)\nonumber\\
  && + \sum_i {M_s\over\bar\rho} U(k_1) 
  \left({m_i\over\bar\rho}\right)^2 u_i(k_2) u_i(k_3)\,U_c(k_1) \nonumber\\
  && + \sum_i\sum_{j\ne i} {m_im_j\over\bar\rho^2} 
        u_i(k_1) u_j(k_2) \,{M_s\over\bar\rho} U(k_3)\,
        U_c(k_1) U_c(k_2) \nonumber\\
  && + \sum_i\sum_{j\ne i}\sum_{k\ne j\ne i} 
    {m_im_jm_k\over\bar\rho^3} u_i(k_1) u_j(k_2)u_k(k_3)
    \nonumber\\
  && \qquad\qquad\qquad\qquad \times \ U_c(k_1) U_c(k_2) U_c(k_3)\nonumber\\
  && + \sum_i\sum_{j\ne i} {m_i\over\bar\rho} u_i(k_1) 
  \left({m_j\over\bar\rho}\right)^2 u_j(k_2) u_j(k_3)\,U_c(k_1)^2 \nonumber\\
  && +\sum_i\left(m_i\over\bar\rho\right)^3 u_i(k_1) u_i(k_2) u_i(k_3)\biggr],
 \label{bk}
\end{eqnarray}
where ${\bm k_3} = {- \bm k_1 - k_2}$.  
Further integrations over appropriate window functions yield the 
spatially smoothed skewness (see, e.g., Scoccimarro et al. 2001; 
Takada \& Jain 2002; Cooray \& Sheth 2002).  Similar relations hold 
for the higher order correlations.  

In the same small mass and size limits which we used when studying 
$\xi(r)$ and $P(k)$, the bispectrum is dominated by the first and last 
terms of equation~(\ref{bk}).  

\section{Averaging over numbers and masses}\label{averages}
So far we have assumed that the masses and numbers of subclumps were 
fixed.  This section shows the result of allowing for a distribution 
of masses and numbers.  Expressions for the power spectra are much 
simpler than for $\xi(r)$, so we will only consider $P(k)$ here.  
Analogously, the averaged bispectrum $B$ is much simpler to evaluate 
than $\xi_{123}$.  
For what is to follow, it is useful to write explicitly that 
the quantity in the previous section is computed at fixed values of 
$M$, ${\cal N}$ and ${\bm m}\equiv \{m_1,\cdots,m_{\cal N}\}$, 
where the masses and density profiles of the ${\cal N}$ subclumps 
may be different.  The quantity we are after is 
\begin{equation}
 \int dM\,{dN(M)\over dM}\,\sum p({\cal N}|M)
                         \,p({\bm m}|{\cal N},M)\,P(k|{\bm m},{\cal N},M),
\end{equation}
where $dN/dM$ is the number density of parent $M$ halos (usually 
called the universal halo mass function), 
$p({\cal N}|M)$ is the probability an $M$ halo has ${\cal N}$ subclumps,  
$p({\bm m}|{\cal N},M)$ is the probability that the ${\cal N}$ 
subclumps were $\{m_1,\cdots,m_{\cal N}\}$ given that there were 
${\cal N}$ of them in the $M$ halo, 
$P(k|{\bm m},{\cal N},M)$ is the quantity we computed in the previous 
section, and the sum is over all values of ${\cal N}$ and ${\bm m}$.  
To proceed, we need models of these different distributions.  

If all the subclumps are identical, and there are ${\cal N}$ of them, 
then each of the terms in the sums over $i$ and $j$ are the same, so the
contribution to the power from the smooth--subclump cross-terms is 
${\cal N}$ times the contribution from a single clump, and the 
contribution from the subclump--subclump terms is 
$\propto {\cal N}({\cal N}-1)\,U_c^2$.  In this case, the total power 
depends on the first two factorial moments of the $p({\cal N}|M)$ 
distribution.  If the subclumps are much smaller than their parents, 
then equation~(\ref{powerapprox}) implies that 
\begin{eqnarray}
P(k) & \approx & \bar n \left(M\over \bar\rho\right)^2 
\left[1-{\cal N}(m/M)^2\right]\ U(k)^2 \nonumber \\
             && + \  {\cal N}\bar n \left(m\over \bar\rho\right)^2 u(k)^2.
\end{eqnarray}
The factor ${\cal N}\bar n$ is simply the number density of the 
subclumps, so it may be useful to think of the two terms above 
separately.  The first term requires knowledge of the mass function of 
the parent halos, whereas the second requires the `mass function' of 
the subclumps.  
As discussed in Section~\ref{details}, accurate formulae for the parent 
mass function are available; subclump mass functions are only just 
becoming available.  

In practice, halos of the same $M$ have different substructure 
distributions.  For example, we expect that there will be some scatter 
around the mean number of subclumps 
$N_{cl}\equiv\langle {\cal N}|M\rangle$ in an $M-$halo, as well as in 
the fraction of mass associated with the clumped component:  
$M_{cl}/M$.  However, if the scatter around the typical subclump 
configuration is small, then we should get a reasonable estimate of 
the effects of substructure if we use a good model of the typical 
configuration, and neglect the fact that there is actually some 
scatter around it.  This is our strategy.  

Let $dn(m|M)/dm$ denote the typical subclump mass function.  Then 
\begin{eqnarray}
 P(k) &=& \int dM\,{dN(M)\over dM}\,
          \left({M_s\over\rho}\right)^2\,U(k|M)^2 \nonumber\\ 
      &+&  2\ \int dM\,{dN(M)\over dM}\,{M_s\over\rho}\,U(k) \nonumber\\ 
      && \qquad\times 
         \int dm\,{dn(m|M)\over dm}\,{m\over\rho}\,u(k|m)\,U_c(k|m)\nonumber\\
      &+& \!\! \int dM\,{dN(M)\over dM}
         \int dm_1\,{dn(m_1|M)\over dm_1}{m_1\over\rho}\,u(k|m_1,M)\nonumber\\
      &&\qquad \times \int dm_2\,{dn(m_2|M,m_1)\over dm_2}\,
                  {m_2\over\rho}\,\nonumber\\
      && \qquad\qquad\times u(k|m_2,M,m_1)\, U_c(k|m_1,m_2)^2\nonumber\\
      &+& \int dM\,{dN(M)\over dM}\nonumber\\
         && \qquad \times dm\,{dn(m|M)\over dm}\,{m^2\over\rho^2}\,u(k|m,M)^2 .
\end{eqnarray}
The third term is the tricky one:  it requires the joint distribution 
of $m_1$- and $m_2$-subclumps in an $M-$halo.  Following Sheth \& 
Lemson (1999), we have chosen to write this as the typical number of 
$m_1$-subclumps in an $M-$halo times the typical number of 
$m_2$-subclumps in $M$-halos which have an $m_1$-subclump.  
Similarly, the notation $U_c(k|m_1,m_2)$ is intended to allow for 
the possibility that the distribution of $m_2$-subclumps within 
the parent $M$-halo may depend on whether or not the parent contains 
an $m_1$-subclump.  

In the limit in which the subclumps are much smaller than the parent, 
and contribute a small fraction of the total mass, the expression 
above becomes 
\begin{eqnarray*}
 P(k) &\approx& \int dM\,{dN(M)\over dM}\,
                \left({M_s\over\rho}\right)^2\,U(k|M)^2 \nonumber\\ 
      &+& \!\! \int dM\,{dN(M)\over dM}\,
         \int dm\,{dn(m|M)\over dm}\,{m^2\over\rho^2}\,u(k|m,M)^2 .
\end{eqnarray*}
The integral over $m$ is from zero to $M$.  If $u(k|m,M)$ is only a 
function of $m$ (i.e., if the density profile of an $m$ subclump 
is independent of the mass of the parent in which it sits), then the 
order of the integrals above can be rearranged.  If we define the 
mass function of the subclumps as 
\begin{equation}
 {dN_c(m)\over dm}\equiv\int_m^\infty dM\,{dN(M)\over dM}\,{dn(m|M)\over dm},
\end{equation}
then 
\begin{eqnarray}
 P(k) &\approx& \int dM\,{dN(M)\over dM}\,
                \left({M_s\over\rho}\right)^2\,U(k|M)^2 \nonumber\\ 
 &&+ \int dm\,{dN_c(m)\over dm}\,\left({m\over\rho}\right)^2\,u(k|m)^2.
\end{eqnarray}
In this limit, the total power is well approximated by adding to the 
power associated with the smooth parent halos the contribution which 
comes integrating over the subclump mass function.  

\section{Details}\label{details}
The results presented above are general.  When used to model 
large-scale structure simulations, one will almost always use 
models of the halo density profiles and halo mass functions 
which have been found to provide good descriptions of numerical 
simulations.  Specifically, the density profiles $F$ and $f$ are 
usually described using the functional form given by 
Navarro, Frenk \& White (1997), 
\begin{equation}
 {\rho(r)\over \bar\rho} = {\Delta_{\rm vir}\over 3\Omega} 
                           {c^3f(c)\over (r/r_s)(1 + r/r_s)^2}
\end{equation}
where the profile is truncated at the virial radius $r_{\rm vir}$, 
$c\equiv r_{\rm vir}/r_s$ is called the concentration, and 
$f(c) = 1/[\ln(1+c)-c/(1+c)]$.  The virial radius $r_{\rm vir}$ 
is defined by requiring that 
$m = 4\pi r_{\rm vir}^3\bar\rho\,\Delta_{\rm vir}$.  
For spatially flat universes with $\Omega_0=(1,0.3)$ and 
$\Lambda=1-\Omega$, $\Delta_{\rm vir}= (178,340)$.
The Fourier transform of the density run around such a halo of 
mass $m$ is 
\begin{eqnarray}
 u(k|m) &=& f(c) \ 
 \Big[ \sin\kappa \Big( {\rm Si}[\kappa(1+c)]- {\rm Si}(\kappa)\Big)
  - \frac{\sin (\kappa c)}{\kappa (1+c)}\nonumber\\
 && \qquad\ +\ \cos \kappa \Big( {\rm Ci}[\kappa(1+c)]-{\rm Ci}(\kappa) 
 \Big)\Big] 
\label{unfw}
\end{eqnarray}
(Scoccimarro et al. 2001), where  $\kappa \equiv kr_{\rm vir}/c$, 
${\rm Si}(x) = \int_0^x {\rm d}t\,\sin (t)/t$ is the sine integral and 
${\rm Ci}(x) = - \int_x^\infty {\rm d}t\,\cos (t)/t$ is the cosine integral 
function.  The concentration parameter of the halos depends on halo 
mass;  we use the parametrization of this dependence given by 
Bullock et al. (2001):  
\begin{equation}
 c(m) \approx {9\over 1+z} \left( {m\over m_*(z)} \right)^{-0.1}.
\end{equation}

Tha parent halo mass function is well described by 
\begin{equation}
 {M^2\over\rho}{dN(M,z)\over dM}\,{{\rm d}M\over M} =
                       \nu f(\nu)\,{{\rm d}\nu\over\nu}, 
\end{equation}
where $\nu\equiv {\delta^2_{\rm sc}(z)/\sigma^2(M)}$, $\rho$ is 
the background mass density, and 
\begin{equation}
 \nu f(\nu) = A(p)\left(1 + (q\nu)^{-p}\right) 
     \,\left({q\nu\over 2\pi}\right)^{1/2}\,\exp\left(-{q\nu\over 2}\right), 
\label{nmfv}
\end{equation}
with $p\approx 0.3$,
$A(p) = [1 + 2^{-p}\Gamma(1/2-p)/\sqrt{\pi}]^{-1} \approx 0.3222$, 
and $q\approx 0.75$ (Sheth \& Tormen 1999).  
Here $\delta_{\rm sc}(z)$ is the critical density required for 
spherical collapse at $z$, extrapolated to the present time using 
linear theory, and 
\begin{equation}
\sigma^2(M) = {4\pi\over (2\pi)^3}\int_0^\infty 
{{\rm d}k\over k}\,k^3P_{\rm Lin}(k)\ W^2(kR_0),
\end{equation}
where $W(x)=(3/x^3)[\sin(x) - x\cos(x)]$ and $R_0=(3M/4\pi\rho)^{1/3}$.  
That is to say, $\sigma(M)$ is the rms value of the initial fluctuation 
field when it is smoothed with a tophat filter of comoving size $R_0$, 
extrapolated using linear theory to the present time.  If $p=1/2$ and 
$q=1$, then $dN/dM$ is the same as the universal mass function 
first written down by Press \& Schechter (1974).  
Note that the mass function is normalized so that 
\begin{equation}
 \int dM\,M\,dN(M)/dM = \rho.
\end{equation}

The subclump distribution is less well constrained.  
Therefore, we will consider two models.  
The first is motivated by the results of numerical simulations 
(Tormen et al. 1998; Ghigna et al. 1999) which suggest that, 
on average, a power law in mass is a reasonable model of the 
typical subclump mass function:  
\begin{equation}
  {dn(m|M)\over dm}\,dm = N_0\left(M\over m\right)^\mu{dm\over m}
                             \qquad {\rm where} \qquad \mu < 1,
\end{equation}
and $N_0$ is a normalization constant which is set by the fraction 
of mass in subclumps:
\begin{displaymath}
 {M_{cl}\over M} = \int_0^M dm\ {m\over M}\,{dn(m|M)\over dm} 
                 = {N_0\over 1-\mu}.
\end{displaymath}
In this model, although the total number of subclumps may diverge, 
the mass contained in them does not.  The mean square subclump mass, 
which is related to the total number of subclump pairs is 
\begin{displaymath}
 M^2\,\int_0^M dm\ \left({m\over M}\right)^2\,{dn(m|M)\over dm} 
    = M^2\,{N_0\over 2-\mu} = M_{cl}M\,{1-\mu\over 2-\mu}.
\end{displaymath}
Simulations suggest $\mu\approx 0.9$ and $M_{cl}/M\approx 0.1$.  
If $P_{\rm Lin}(k)\propto k^n$ and we use the Press--Schechter form 
for $dN/dM$, then in this power law subclump model, the subclump mass 
function is 
\begin{eqnarray}
 {m^2\over\rho}{dN_c(m)\over dm} &=& {2^{(\mu-1)/\alpha}\over\Gamma(1/2)}\,
                         \left({M_*\over m}\right)^{\mu-1}\nonumber\\
         && \qquad \times \ \Gamma\left({1\over 2} +
                          {\mu-1\over\alpha},{(m/M_*)^\alpha\over 2}\right),
\end{eqnarray}
where $\alpha=(n+3)/3$ (the value $n\approx -3/2$ is a reasonable 
approximation to most CDM models).  At small masses, $m/M_*\ll 1$, 
the subclump mass function is a power law:  $dN_c/dm\propto m^{-\mu-1}$, 
whereas at $m/M_*\gg 1$ it drops exponentially.  

In practice, the simulations only have finite mass resolution, so a 
power law model with a minimum mass cutoff may be more appropriate.  
A low-mass cutoff may also be a useful model if one is only interested 
in subclumps which could host galaxies.  In this case, the mass 
fraction in subclumps is 
\begin{displaymath}
 {M_{cl}\over M} = \int_{M_{min}}^M dm\ {m\over M}\,{dn(m|M)\over dm} 
                 = N_0\,{1 - (M_{min}/M)^{1-\mu}\over 1-\mu}.
\end{displaymath}
The lower mass cutoff means that the number of subclumps no longer 
diverges:
\begin{displaymath}
 N_{cl} = \int_{M_{min}}^M dm\ {dn(m|M)\over dm} 
                 = N_0\,{(M/M_{min})^\mu - 1\over \mu} .
\end{displaymath}
Therefore, when $M_{min}\ll M$, the first term is the dominant one, 
and the number of subclumps increases as a power law in $M$.
In practice, $\mu$ is sufficiently close to unity that the mass fraction 
in subclumps is approximately $N_0/(1-\mu)$ only for the most massive 
halos.  

Our second model of $dn(m|M)/dm$ is to assume that it is given by the 
progenitor distribution at some higher redshift $z_1$, and all 
progenitors lost the same fraction of mass as they fell into the parent 
they now occupy, so they are all a fraction $f$ of the mass they were 
at $z_1$.  In this case, 
\begin{equation}
 {dN_c(m,z_0)\over dm} = \int_{m/f}^\infty \!\!\!\!\!\!
                          dM\ {dN(M,z_0)\over dM}\,{dn(m/f,z_1|M,z_0)\over dm}.
\end{equation}
Mass conservation means that this integral reduces to the mass function 
at the higher redshift:  $dN_c(m,z_0)/dm = dN(m/f,z_1)/dm$.  
The total amount of mass in these subclumps is 
\begin{displaymath}
 \int dm\,m\,dN_c(m)/dm = f\,\int dm\,(m/f)\,dN(m/f)/dm = f\rho.
\end{displaymath}
This provides a convenient way of thinking about the subclump mass 
function.  In practice, it is likely that the subclumps which have 
survived to the present come from a distribution of earlier epochs, 
rather than a single epoch $z_1$, and the distribution of $z_1$ might 
peak at different redshifts for different values of $m/M$.  Ignoring 
this subtlety, which is what the model above does, should be a 
reasonable approximation if the distribution around the mean $z_1$ is 
not broad.  In practice, for small values of $m/M$, the subclump mass 
function is reasonably well approximated by a power law, and so this 
model also gives a subclump mass function which is in reasonable 
agreement with the simulations.  

The virtue of this model is that it allows a simple estimate of the 
shape of the correlation function in the limit in which the subclumps 
are each a small fraction of the mass of the parent halo, and the 
total mass in subclumps is a small fraction of the total mass.  Recall 
that, in this limit, the total correlation function is well approximated 
by the sum of two terms.  The first term is the contribution from pairs 
which are from the smoothly distributed component, and the second is 
from the subclumps.  The subclump contribution, then, can be determined 
by rescaling the contribution from the smooth component at $z_1$.  
If the particles which were stripped from the halos present at $z_1$ 
were random particles, then the number of pairs is lower by a factor 
of $f^2$; since the halos at $z_1$ were a factor of $(1+z_1)^3$
denser, the contribution to the correlation function is different 
by a factor of $f^2 (1+z_1)^3$.  This yields a factor which is probably 
slightly smaller than unity.  On the other hand, if the mass was stripped 
from the earlier halos in shells, much like layers off an onion, as 
simulations suggest, then a better model of the subclump contribution 
is to truncate the halos which were present at $z_1$ at a fraction 
$\sim f$ of their virial radii when estimating how the number of 
pairs changes with scale.  Although this changes the actual shape of 
the correlation function (e.g., the subclump pairs are shifted to 
scales which are a factor of $f$ smaller), the typical factor by which 
the correlations are affected is $f^{-1} (1+z_1)^3$, which can be 
considerably larger than unity.  

\section{Discussion}\label{discuss}

We have shown how to incorporate the effects of substructure into the 
halo model description of the nonlinear density field.  Accounting 
for this substructure is important on scales smaller than the virial 
radii of typical halos.  The effects are more pronounced for statistics 
which treat the subclumps preferentially, such as the power spectrum 
measured in studies of weak galaxy--galaxy gravitational lensing.  
Substructure will also change the dynamics within halos.  
Although we have not done so here, it is straightforward to insert 
our model for substructure into the halo model of the cosmic virial 
theorem, and the mean pairwise velocity and velocity dispersion 
developed in Sheth et al. (2001).

The stable clustering limit is a physically appealing description of 
clustering on small scales (Peebles 1980).  It has been argued that
a model with smooth halos 
is inconsistent with this limit (Ma \& Fry 2000; Scoccimarro et al. 2001).  
Substructure changes the shape of the small scale power spectrum 
(c.f., Fig.~\ref{pksubclumps}); at least in principle, it can bring 
the halo model predictions into agreement with the stable clustering 
solution.  However, it is not obvious that stable clustering is, 
indeed, the correct physical limit.  Smith et al. (2002) argue that 
the stable clustering assumption is inconsistent with the results of 
high resolution numerical simulations.  They also find that the 
simulations do not follow the small scale scaling predicted by models 
in which halos are smooth.  Once an accurate model of the subclump 
mass function is available, it will be interesting to compare the 
predictions of our description of substructure with their results.  

Although we have focussed primarily on the implications of substructure 
for the halo model of nonlinear clustering, our results have a wide 
range of other applications.  For example, excess power in the Fourier 
transforms of images of galaxies or distant clusters can be used to 
infer the existence of spiral arms or substructure.  This is the 
subject of work in progress.  Closely related is the question of what 
images of high redshift galaxies may look like.  Observations through 
a filter which has a fixed wavelength range probe the emission from 
high redshift galaxies at shorter restframe wavelengths than for 
galaxies at low redshift.  
If obscuration by dust is not a problem, and the UV luminosity is 
dominated by patchy star forming regions, then the images of high 
redshift galaxies should show considerable substructure.  Our results 
suggest that, in this case, the power spectrum obtained by Fourier 
transforming the image of a high redshift patch of sky should show an 
increase in small scale power.  

In addition, although we have phrased the entire discussion of 
substrucutre in terms of spatial statistics, this is not really 
necessary.  Large databases describing various observed 
characteristics of galaxies are now becoming available 
(e.g., the 2dFGRS and SDSS surveys).  If some of $n$ observables are 
correlated with others, the data will not fill the full 
$n-$dimensional space available:  the data set itself can be thought 
of as being clumpy, and the various clumps in dataspace may 
themselves have substructure.  The formalism developed here provides 
a way of discovering, quantifying and modelling such substructure.  

\bigskip

\noindent We thank Masahiro Takada for helpful discussions. B.J. acknowledges
financial support from a NASA-LTSA grant and a Keck foundation grant.

\end{document}